\documentclass{aa}  

\usepackage{graphicx}
\usepackage{txfonts}
\usepackage[breaklinks,colorlinks=true,citecolor=black,linkcolor=black,urlcolor=blue,bookmarks=true]{hyperref}
\usepackage{threeparttablex}

\usepackage{pgfplots}
\pgfplotsset{compat=1.17}
\usepackage{mathrsfs}
\usetikzlibrary{arrows}
\definecolor{rvwvcq}{rgb}{0.08235294117647059,0.396078431372549,0.7529411764705882}

\newcommand{\kms}{\mbox{km\,s$^{-1}$}}
\newcommand{\kmsMpc}{\kms{}Mpc$^{-1}$}
\newcommand{\LCDM}{$\Lambda$CDM}
\newcommand{\Vh}{v_\mathrm{h}}
\newcommand{\Vg}{v_\mathrm{G}}
\newcommand{\Msun}{\mbox{$M_\sun$}}

\begin{document} 

\title{The frozen outskirts: a cold Hubble flow and the mass of the Local Group}

\author{
Danila Makarov \inst{1}
\and
Dmitry Makarov \inst{1}\fnmsep\thanks{\email{dim@sao.ru}}
\and
Lidia Makarova \inst{1}
\and
Noam Libeskind \inst{2}
}

\institute{
Special Astrophysical Observatory, Russian Academy of Sciences, Nizhnii Arkhyz, 369167 Russia
\and
Leibniz Institut f\"{u}r Astrophysik Potsdam (AIP), An der Sternwarte 16, D-14482, Potsdam, Germany
}

\date{}

 
\abstract{
We analyze the velocity field of peripheral members of the Local Group.
The Hubble flow at distances from 400 to 1400~kpc, formed by 7 of 11 nearby galaxies, is characterized by an extremely small line-of-sight velocity dispersion of 15~\kms{}, which differs significantly from the predictions of cosmological simulations of about 70~\kms{}.
This fact allows us to determine the total mass of the Local Group as $M_\mathrm{LG} = (2.47 \pm 0.15) \times 10^{12}$~\Msun{} using an analytical model of the Hubble flow around a spherical overdensity in the standard flat \LCDM{} universe.
The practical equality of this mass to the sum of the masses of our Galaxy and the Andromeda Galaxy, as well as the absence of mass growth in the range of distances under consideration, gives grounds to conclude that the entire mass of the Local Group is confined within the virial radii around its two main galaxies.
The barycenter, found from the minimal scatter of mass estimates, corresponds to the mass ratio of the Milky Way and the Andromeda Galaxy equal to $M_\mathrm{MW}/M_\mathrm{M31} = 0.74\pm0.10$.
The velocity of our Galaxy to the barycenter turns out to be $62.6\pm2.6$~\kms{}.
This allows us to determine the apex of the Sun relative to the barycenter of the Local Group to be $(l,b,V) = ( +94.0^\circ \pm 0.7^\circ ,  -2.7^\circ \pm 0.3^\circ , 301 \pm 3~\kms{})$ in the Galactic coordinates.
}


\maketitle
%

\section{Introduction}

Analysis of the kinematics of satellites within virial zones is the primary method for estimating the mass of galactic systems, using various analogues of the virial theorem \citep{1985ApJ...298....8H}.
In the case of distant systems, we can only measure the line-of-sight velocities and the angular separation between galaxies.
Thanks to high-precision distance measurements, we know the three-dimensional distribution of satellites around the Andromeda Galaxy (M\,31), which allows us to improve estimates of its mass.
Due to the Gaia mission~\citep{2018A&A...616A...1G}, proper motions have been measured for most of the Milky Way (MW) satellites, unambiguously determining all six components of phase space for each satellite.
This information is actively used to clarify the structure of our Galaxy and estimate its total mass.

Dwarf galaxies located outside the virial zones of the MW and M\,31 provide an alternative method for estimating the total mass of the Local Group by the deviation of their velocities from the linear Hubble law.
Analysis of the peculiar motions of galaxies is a key technique for mapping the distribution of matter on scales from a few to hundreds of megaparsecs.
As an example, \citet{2014Natur.513...71T} discovered the Laniakea supercluster, which includes the Local Supercluster, by coherent flow of galaxies into this region.
The measurement of the zero-velocity sphere radius, also known as the stopping or turning radius, has become the standard method for the total mass estimation of the Local Group~\citep{2009MNRAS.393.1265K} and other nearby galaxy groups~\citep{2018A&A...609A..11K}.
It should be emphasized that inside the stopping radius, falling objects are in a nonlinear regime and it is necessary to use more complex models to describe their motion.
The exact solution for the Hubble flow~\citep{2020PhRvD.102h3529B} for a spherically symmetric distribution of matter gives us a tool to estimate the mass at different distances from the gravitating center.

\begin{figure}
\centering
\includegraphics[width=\linewidth, clip]{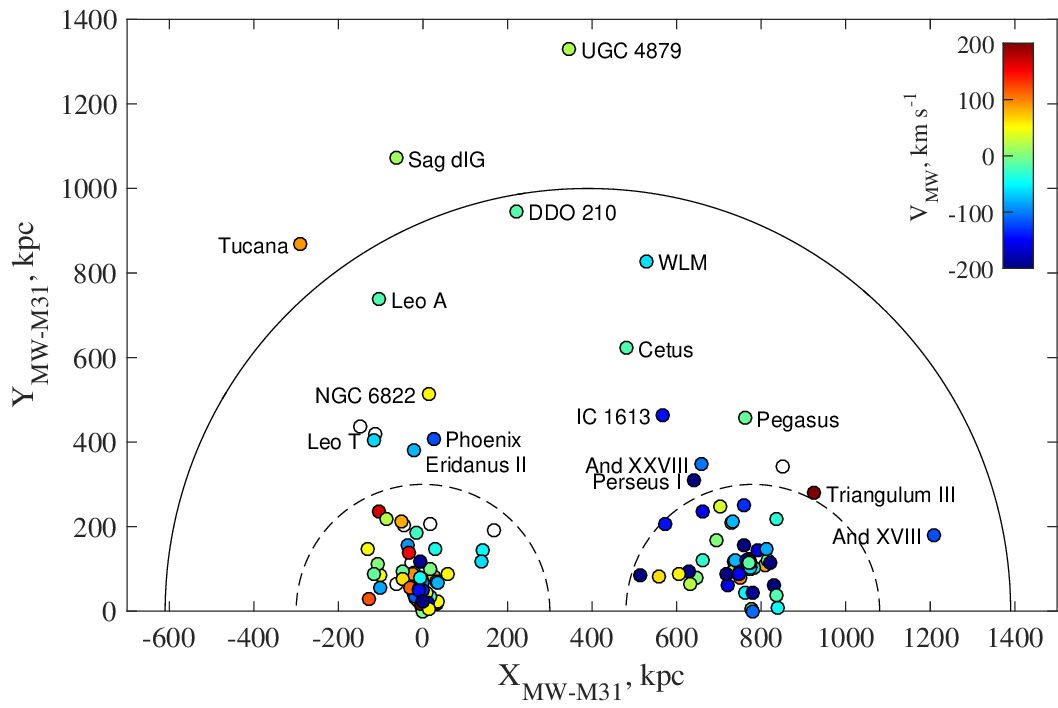}
\caption{
Distribution of Local Group galaxies in a cylindrical projection, where the X axis is directed from the center of the Milky Way to the center of the Andromeda Galaxy, and the Y axis corresponds to the distance from the X axis.
The dashed semicircles with a radius of 300~kpc roughly correspond to the size of the virial zones around the two main galaxies.
A solid semicircle with a radius of 1~Mpc shows the approximate boundary of the Local Group.
Objects are colored according to their radial velocities relative to the center of our Galaxy.
Open dots indicate dwarf galaxies with unknown velocities.
}
\label{fig:LG}
\end{figure}

The structure of the Local Group is more complex than this simple model.
It exhibits a well-known dumbbell-shaped distribution around two giant galaxies of comparable mass, the MW and M\,31 (see Fig.~\ref{fig:LG}).
All other members of the Local Group have substantially lower masses and luminosities.
To date, 134 galaxies have been identified within 1~Mpc of the Local Group barycenter\footnote{
This total of 134 galaxies includes the MW and M\,31 groups, as compiled by \citet{2025arXiv250312612M}, along with additional peripheral members of the Local Group, assembled in this work (see Table~\ref{tab:GalaxyList}).
}.
Most of them are concentrated in the virial zones around the MW and M\,31.
There are 67 satellites of the MW within 300~kpc and 48 satellites of the M\,31 within the same 300~kpc range~\citep{2025arXiv250312612M}.
Figure~\ref{fig:LG} illustrates that dwarf galaxies outside the virial zones are concentrated along planes passing through the two main galaxies, oriented roughly perpendicular to the MW\,--\,M\,31 line.

Despite this, we can assume that the deviations of the Local Group's potential from spherical symmetry are relatively small at its periphery. 
Therefore, considering the motions of galaxies relative to the center of mass of the MW and M\,31, we can expect that the infall of galaxies into the Local Group is governed by the total mass of the system and directed toward its barycenter.

The aim of the work is to analyze the motion of galaxies outside the virial zones, estimate the total mass of the Local Group based on the shape of the Hubble flow and attempt to measure changes in the mass profile at the periphery of the system.

\section{Velocity field model}

Despite the fact that proper motions have been measured for a large number of even distant members of the Local Group, these data are of little use for our analysis.
For galaxies at distances greater than 300~kpc, errors range from 0.01 to 0.4~mas\,yr$^{-1}$, with an average value of 0.14~mas\,yr$^{-1}$~\citep{2022A&A...657A..54B}, which translates into typical velocity errors at such distances of hundreds of \kms{}.
The proper motion is found to be significant at a level greater than 3 sigma for only two galaxies, NGC\,6822 and WLM.
However, even in these cases, the tangential velocity errors are approximately 35 and 170~\kms{} for NGC\,6822 and WLM, respectively, which are significantly worse than the typical accuracy of line-of-sight velocity measurements, which is of the order of several \kms{}.
That is why \citet{2022A&A...657A..54B} caution the reader that for distant galaxies ``the error in transverse velocity is still too large for scientific applications.''
Thus, all further analysis is focused exclusively on the line-of-sight velocities of galaxies.

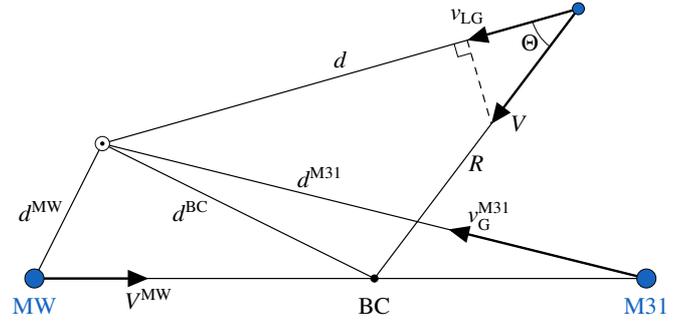
\begin{figure}
\centering
\resizebox{\columnwidth}{!}{
\begin{tikzpicture}[line cap=round,line join=round,>=triangle 45]
\draw [line width=0.5pt] (0,0)--(9,0);  
\draw [line width=0.5pt] (0,0)--(1,2);  
\draw [line width=0.5pt] (1,2)--(5,0);  
\draw [line width=0.5pt] (1,2)--(9,0);  
\draw [line width=0.5pt] (1,2)--(8,4);  
\draw [line width=0.5pt] (5,0)--(8,4);  

\draw [->,line width=1pt] (8,4) -- ({8-(8-1)*1.7/7.28011},{4-(4-2)*1.7/7.28011}); 
\draw [->,line width=1pt] (8,4) -- ({8-(8-5)*1.7/5/0.79669},{4-(4-0)*1.7/5/0.79669}); 
\draw [dashed,line width=0.5pt] ({8-(8-1)*1.7/7.28011},{4-(4-2)*1.7/7.28011}) -- ({8-(8-5)*1.7/5/0.79669},{4-(4-0)*1.7/5/0.79669}) ;
\draw [line width=0.2pt] ({8-(8-1)*1.9/7.28011},{4-(4-2)*1.9/7.28011}) -- ({8-(8-1)*1.9/7.28011+2*0.2/7.28011},{4-(4-2)*1.9/7.28011-7*0.2/7.28011}) -- ({8-(8-1)*1.9/7.28011+2*0.2/7.28011+7*0.2/7.28011},{4-(4-2)*1.9/7.28011-7*0.2/7.28011+2*0.2/7.28011}) ;

\draw [line width=0.2pt] ({8-0.7*0.96152},{4-0.7*0.27472}) arc ({180+15.94540}:{180+53.13010}:0.7) ;

\draw [->,line width=1pt] (9,0) -- ({9-(9-1)*3/8.24621},{0+(2-0)*3/8.24621}); 
\draw [->,line width=1pt] (0,0) -- (1.67890,0); 

\draw [fill=rvwvcq] (0,0) circle (4pt); 
\draw [fill=rvwvcq] (9,0) circle (4pt); 
\draw [fill=rvwvcq] (8,4) circle (2.5pt); 
\draw [fill=black] (5,0) circle (1.5pt);  
\draw [fill=white] (1,2) circle (3pt);  
\draw [fill=black] (1,2) circle (0.7pt);  

\draw[color=rvwvcq] (0,-0.4) node {MW};
\draw[color=rvwvcq] (9,-0.4) node {M31};
\draw[color=black] (5,-0.4) node {BC};

\draw[color=black] ({0.5-0.4},{1}) node {$d^\mathrm{MW}$};
\draw[color=black] ({4-(4-1)/2-0.2},{1}) node {$d^\mathrm{BC}$};
\draw[color=black] ({1+(9-1)/2.5},{2-(2-0)/2.5+0.3}) node {$d^\mathrm{M31}$};
\draw[color=black] ({1+(8-1)/2},{2+(4-2)/2+0.25}) node {$d$};

\draw[color=black] ({5+(8-5)/2},{0+(4-0)/2-0.3}) node {$R$};
\draw[color=black] ({8-(8-1)*1.7/7.28011},{4-(4-2)*1.7/7.28011+0.35}) node {$v_\mathrm{LG}$};
\draw[color=black] ({8-(8-5)*1.7/5/0.79669+0.4},{4-(4-0)*1.7/5/0.79669}) node {$V$};
\draw[color=black] ({9-(9-1)*3/8.24621+0.6},{0+(2-0)*3/8.24621+0.2}) node {$v^\mathrm{M31}_\mathrm{G}$};
\draw[color=black] (1.67890,-0.3) node {$V^\mathrm{MW}$};
\draw[color=black] (8-0.71,4-0.5) node {$\Theta$};

\end{tikzpicture}
}
\caption{
Scheme for calculating the velocity field in the vicinity of the Local Group.
The symbol $\odot$ marks the location of the observer.
The positions of the MW, M\,31, and the galaxy under study are shown as blue dots. 
The barycenter (BC) of the Local Group is indicated by a black dot. 
Corresponding distances and velocity vectors are labeled as described in the text.
}
\label{fig:Schema}
\end{figure}

Figure~\ref{fig:Schema} presents a schematic diagram of the corrections described below.
The barycenter of the Local Group is designated as BC.
Small letters refer to observed values, and uppercase letters denote values relative to the barycenter.
The blue dots indicate real objects: MW, M\,31, and a galaxy under consideration.
The position of the observer is marked by the Sun symbol.
As part of our Galaxy, the observer does not participate in the Hubble flow independently.

First of all, we take into account the motion of the Sun in the Galaxy, transferring heliocentric velocities, $\Vh$, into the Galactocentric system.
\begin{equation}
\Vg = \Vh + \vec{V_\odot}\cdot\vec{n},
\end{equation}
where $\vec{n}$ is an unit direction vector to a galaxy,
and $\vec{V_\odot}$ is the solar velocity vector of $(V_x,V_y,V_z) = (9.5, 250.7, 8.56)$~\kms{} in Galactic coordinates with respect to the MW center~\citep{2024MNRAS.530..710A}. 
It is based on 18 years of observations of the proper motion of Sgr\,A*~\citep{2020ApJ...892...39R} and the distance of $d^\mathrm{MW}=8249 \pm9 \pm45$~pc to the central supermassive black hole~\citep{2021A&A...647A..59G}.

Next, we assume that the barycenter of the Local Group is primarily determined by its two dominant galaxies and lies along the line connecting the centers of the MW and M\,31.
For a given MW-to-M\,31 mass ratio $x$, the center-of-mass position vector $\vec{d}^\mathrm{BC}$ is determined by the following equation.
\begin{equation}
\vec{d}^\mathrm{BC} = \frac{x}{1+x}\vec{d}^\mathrm{MW} + \frac{1}{1+x}\vec{d}^\mathrm{M31},
\end{equation}
where $\vec{d}^\mathrm{MW}$ and $\vec{d}^\mathrm{M31}$ are the position vectors of the Milky Way and the Andromeda Galaxy, respectively.
We adopt a distance to M\,31 of $d^\mathrm{M31} = 776.2^{+22}_{-21}$~kpc~\citep{2022ApJ...938..101S}, resulting in a separation of 780.3~kpc between the two galaxy centers.

Then, knowing the velocity vector of the Andromeda Galaxy relative to our Galaxy and the mass ratio of these galaxies, we can determine the MW velocity vector relative to the barycenter of the Local Group.
The heliocentric velocity of M\,31 is known with good accuracy, $\Vh^\mathrm{M31}=-301.0\pm1.0$~\kms{}~\citep{2013MNRAS.430..971W}.
After correction for the motion of the Sun, we determine its Galactocentric velocity to be $\Vg^\mathrm{M31}=-109\pm2$~\kms{}.
\citet{2021MNRAS.507.2592S} note that the M\,31 transverse velocity remains poorly determined and conclude that the approach of these two giant galaxies is radial.
In agreement with this conclusion, we also assume that the tangential velocity of the Andromeda Galaxy is zero.
Thus, the motion of our Galaxy to the barycenter of the Local Group is described by the equation.
\begin{equation}
\vec{V}^\mathrm{MW} = \frac{ \Vg^\mathrm{M31} }{1+x} \vec{n}^\mathrm{MW-M31},
\end{equation}
where the unit vector $ \vec{n}^\mathrm{MW-M31} = ( \vec{d}^\mathrm{M31}-\vec{d}^\mathrm{MW} ) / |\vec{d}^\mathrm{M31}-\vec{d}^\mathrm{MW}| $ defines the direction from the center of MW to the center of M\,31.
The appropriate correction to the observed line-of-sight velocity, accounting for the motion of the MW within the Local Group, is
\begin{equation}
v_\mathrm{LG} = \Vg + \vec{V}^\mathrm{MW}\cdot\vec{n},
\end{equation}
where, as before, $\vec{n}$ is the unit vector in the direction of the galaxy.

Then, assuming that the outer members of the Local Group head toward or away from its barycenter, we can determine their barycentric velocities by deprojecting the line-of-sight velocities (after applying all the corrections described above):
\begin{equation}
V = \frac{v_\mathrm{LG}}{\cos\Theta}.
\end{equation}
The distance of a galaxy from the barycenter is determined in the usual manner:
\begin{equation}
\vec{R} = \vec{d}-\vec{d}^\mathrm{BC}.
\end{equation}

The analytical solution derived by \citet{2020PhRvD.102h3529B} for a standard flat \LCDM{} universe, where $\Omega_\Lambda + \Omega_M = 1$, describes the Hubble flow around a mass concentration as a function of the density enclosed within a sphere of a given radius,
and is given in equations (6), (10), and (14) in the original study.
Consequently, based on the velocity $V$ and separation $R$ of a galaxy relative to the barycenter of the Local Group, we can solve the inverse problem to estimate the mass enclosed within radius $R$ and reconstruct the underlying mass distribution.
In our work, we adopt the cosmological parameters of the standard $\Lambda$CDM model: $\Omega_\Lambda=0.685 \pm 0.007$, $\Omega_M=0.315 \pm 0.007$, and $H_0=67.4 \pm 0.5$~\kmsMpc{}, based on the Planck mission data~\citep{2022arXiv220108666L}.

\section{Mass estimation from the Hubble flow model}

\begin{table*}
\centering
\begin{ThreePartTable}
\caption{List of galaxies outside virial zones of MW and M\,31}
\label{tab:GalaxyList}
{\small
\begin{tabular}{lcc@{~~}l@{~}lllr@{~}llrr}
\hline\hline
Name       & J2000.0           & \multicolumn{3}{c}{$(m-M)_0$} & Method &                        & \multicolumn{2}{c}{$\Vh$} &  & $D_\mathrm{MW}$ & $D_\mathrm{M31}$ \\
\cline{3-5}\cline{8-9}\cline{11-12}
           &                   & \multicolumn{3}{c}{mag}   &        &                            & \multicolumn{2}{c}{\kms{}} &  & \multicolumn{2}{c}{kpc}  \\
\hline
WLM            & 000158.1$-$152740 & 24.96 & $+0.07$ & $-0.07$ & TRGB   & $a$ & $-122.0$ & $\pm 2.0$ & $\alpha$   &  981 &  865 \\
And XVIII      & 000214.5$+$450520 & 25.43 & $+0.05$ & $-0.03$ & HB     & $b$ & $-337.2$ & $\pm 1.4$ & $\beta$    & 1222 &  465 \\
Cetus          & 002611.0$-$110240 & 24.48 & $+0.10$ & $-0.10$ & TRGB   & $a$ & $ -83.9$ & $\pm 1.2$ & $\gamma$   &  788 &  691 \\
IC 1613        & 010447.8$+$020800 & 24.32 & $+0.05$ & $-0.05$ & RR Lyr & $c$ & $-231.0$ & $\pm 1.0$ & $\delta$   &  734 &  511 \\
Triangulum III & 012141.3$+$262332 & 24.92 & $+0.07$ & $-0.07$ & TRGB   & $d$ & $ 138.6$ & $\pm 0.5$ & $\epsilon$ &  966 &  315 \\
Phoenix        & 015106.3$-$442641 & 23.06 & $+0.12$ & $-0.12$ & TRGB   & $e$ & $ -21.2$ & $\pm 1.0$ & $\zeta$    &  409 &  859 \\
Perseus I      & 030123.6$+$405918 & 24.24 & $+0.06$ & $-0.06$ & RR Lyr & $c$ & $-325.9$ & $\pm 3.0$ & $\eta$     &  711 &  340 \\
Eridanus II    & 034421.1$-$433159 & 22.80 & $+0.10$ & $-0.10$ & BHB    & $f$ & $  75.6$ & $\pm 1.3$ & $\theta$   &  365 &  880 \\
UGC 4879       & 091602.2$+$525024 & 25.68 & $+0.03$ & $-0.03$ & TRGB   & $a$ & $ -29.2$ & $\pm 1.6$ & $\gamma$   & 1373 & 1398 \\
Leo K          & 092406.1$+$163038 & 23.16 & $+0.14$ & $-0.55$ & CMD    & $g$ &          &           &            &  434 &  986 \\
Leo T          & 093453.4$+$170305 & 23.10 & $+0.10$ & $-0.10$ & TRGB   & $e$ & $  38.1$ & $\pm 2.0$ & $\iota$    &  422 &  984 \\
Leo A          & 095926.4$+$304447 & 24.44 & $+0.11$ & $-0.11$ & RR Lyr & $h$ & $  23.9$ & $\pm 0.1$ & $\iota$    &  777 & 1175 \\
Leo M          & 110521.2$+$252043 & 23.30 & $+0.41$ & $-0.09$ & CMD    & $g$ &          &           &            &  460 & 1025 \\
Sag dIG        & 192959.0$-$174041 & 25.17 & $+0.08$ & $-0.08$ & TRGB   & $e$ & $ -79.0$ & $\pm 1.0$ & $\alpha$   & 1074 & 1365 \\
NGC 6822       & 194457.7$-$144811 & 23.58 & $+0.07$ & $-0.07$ & TRGB   & $a$ & $ -54.5$ & $\pm 1.7$ & $\gamma$   &  513 &  923 \\
DDO 210        & 204651.8$-$125053 & 24.95 & $+0.14$ & $-0.14$ & TRGB   & $a$ & $-139.5$ & $\pm 2.0$ & $\kappa$   &  972 & 1099 \\
And XXVIII     & 223241.2$+$311258 & 24.36 & $+0.05$ & $-0.05$ & RR Lyr & $c$ & $-331.1$ & $\pm 1.8$ & $\lambda$  &  745 &  368 \\
Tucana         & 224149.0$-$642512 & 24.82 & $+0.03$ & $-0.03$ & TRGB   & $a$ & $ 194.0$ & $\pm 4.3$ & $\mu$      &  916 & 1378 \\
Pegasus        & 232834.1$+$144448 & 24.74 & $+0.05$ & $-0.05$ & RR Lyr & $c$ & $-184.5$ & $\pm 0.3$ & $\nu$      &  888 &  458 \\
Pegasus W      & 235315.0$+$220607 & 24.81 & $+0.14$ & $-0.22$ & CMD    & $i$ &          &           &            &  918 &  349 \\
\hline\hline
\end{tabular}
\begin{tablenotes}
\item\textbf{Distance references:}
$a$) \cite{2021AJ....162...80A};
$b$) \cite{2019MNRAS.489..763W};
$c$) \cite{2022ApJ...938..101S};
$d$) \cite{2024MNRAS.528.2614C};
$e$) \cite{2020AJ....160..124M};
$f$) \cite{2016ApJ...824L..14C};
$g$) \cite{2024ApJ...967..161M};
$h$) \cite{2013MNRAS.432.3047B};
$i$) \cite{2023ApJ...944...14M}.
\item\textbf{Velocity references:}
$\alpha$)   \cite{2004AJ....128...16K};
$\beta$)    \cite{2024ApJ...972..180K};
$\gamma$)   \cite{2014MNRAS.439.1015K};
$\delta$)   \cite{2018ApJ...861...49H};
$\epsilon$) \cite{2015ApJS..219...12A};
$\zeta$)    \cite{2017MNRAS.466.2006K};
$\eta$)     \cite{2014ApJ...793L..14M};
$\theta$)   \cite{2022ApJ...940..136P};
$\iota$)    \cite{2020AJ....160..124M};
$\kappa$)   \cite{2006MNRAS.365.1220B};
$\lambda$)  \cite{2013ApJ...768...50T};
$\mu$)      \cite{2009A&A...499..121F}.
$\nu$)      \cite{2013MNRAS.430..971W};
\end{tablenotes}
}
\end{ThreePartTable}
\end{table*}

\begin{figure}
\centering
\includegraphics[width=\linewidth, clip]{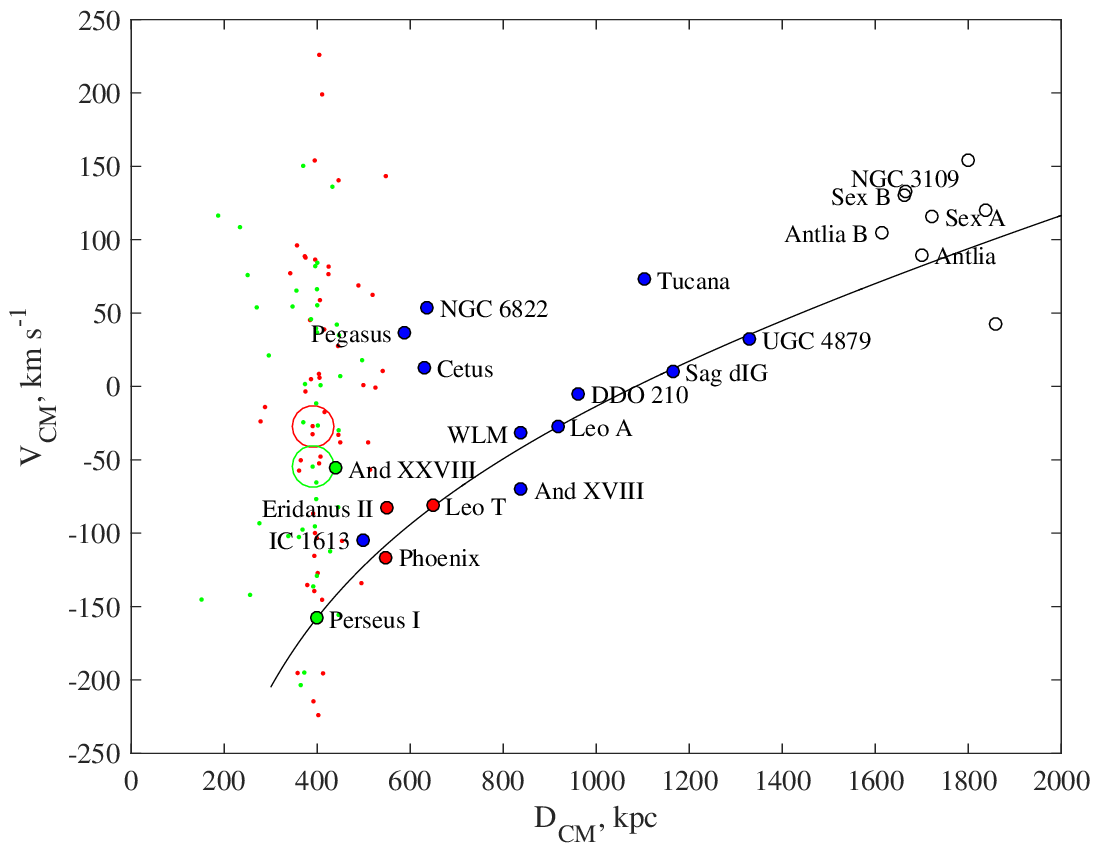}
\caption{
The Hubble flow with respect to the Local Group center of mass.
The big red open circle corresponds to the MW, the small red dots indicate its satellites inside 300~kpc and the filled red circles represent the MW members at distances from 300 to 450~kpc.
Likewise, the M\,31 and its satellites are shown in green.
The galaxies selected for the analysis are shown in blue.
The solid line corresponds to the Hubble flow model around a point mass of $2.5\times10^{12}$~$\Msun$.
}
\label{fig:HubbleFlow}
\end{figure}

The Hubble flow as it would look from the Local Group centroid, located strictly midway between the MW and M\,31 is shown in Fig.~\ref{fig:HubbleFlow}.
There is a noticeable gap between UGC\,4879 at 1.35~Mpc and the next ``cloud'' of objects beyond 1.6~Mpc, which includes Antlia\,B, Sextants\,B, NGC\,3109, Antlia, Sextans\,A and so on.
This provides us a natural upper limit for the use of the model, as the analysis of the motion of more distant galaxies may need to consider the influence of neighboring giant galaxies---M\,81, M\,94, Cen\,A, NGC\,253, and IC\,342---located at distances of 3--4~Mpc.

To date, there are 20 known peripheral Local Group members at distances closer than 1400~kpc and beyond 300~kpc around the Milky Way and the Andromeda Galaxy.
Their list is given in Table~\ref{tab:GalaxyList}.
It presents the names, J2000.0 coordinates, distance moduli with errors, methods and sources of the data, observed heliocentric velocities with errors, and their sources.
The last two columns provide the distance from the centers of the MW and M\,31, respectively.
High-precision distances for all of these galaxies are known, while the radial velocities for three of them have not yet been measured. 

As can be seen in Fig.~\ref{fig:HubbleFlow}, the motion of distant members of the Local Group is fairly regular.
With the exception of six galaxies (And\,XXVIII, Triangulum\,III\footnote{Triangulum\,III is not shown in Fig.~\ref{fig:HubbleFlow} due to its extremely high peculiar velocity of $+333$~\kms{} relative to the Local Group centroid.}, Pegasus, Cetus, NGC\,6822, and Tucana), the velocity field outside the 300~kpc virial zones around the MW and M\,31 is reasonably well described by a simple model of the Hubble flow~\citep{2020PhRvD.102h3529B} around a point mass.

We have marked objects lying in the 300 to 450~kpc layer from the MW in red (Phoenix, Eridanus\,II and Leo\,T), and similarly galaxies around M\,31 in green (Perseus\,I, Triangulum\,III and And\,XXVIII).
In fact, they are located near the virial zones and should be dominated by the nearest giant galaxy, and consequently our simple model of the infall on the center of mass of the Local Group may not work.
Moreover, Triangulum\,III is a satellite of M\,33~\citep{2024MNRAS.528.2614C}.
Therefore, it cannot be considered as a free-falling particle.
Thus, we do not include these six galaxies in the analysis.
Nevertheless, it should be noted that four of them also follow the model quite well.

As a result, our final sample contains 11 galaxies located within 1400~kpc of the Local Group barycenter and at distances greater than 450~kpc from the MW and M\,31.

As a first guess, we can determine the position of the barycenter of the Local Group based on the individual masses of the MW and M\,31.
Estimates of the total mass estimates of our Galaxy vary widely, ranging from 0.4 to $1.6\times10^{12}$~\Msun. 
Using 33 measurements within 200~kpc compiled by \citet{2020SCPMA..6309801W}, the average is $(1.07\pm0.05)\times10^{12}$~\Msun{} with a standard deviation of $0.3\times10^{12}$~\Msun{}.
\citet{2023ARep...67..812B} obtain a slightly lower value of $(0.88\pm0.06)\times10^{12}$~\Msun{} with a standard deviation of $0.24\times 10^{12}$~\Msun{} using 20 recent studies.
As a rule, mass measurements of the Andromeda Galaxy lead to values 1.5--2 times higher than the MW mass.
The average of 7 recent estimates within 300~kpc collected by \citet{2023arXiv230503293B} is $(1.61 \pm 0.09) \times 10^{12}$~\Msun{} with the standard deviation of $0.24 \times 10^{12}$~\Msun{}.

Using a modification of the projection mass method \citep{1981ApJ...244..805B}, which accounts for the three-dimensional distribution of galaxies in the group and assumes an isotropic distribution of satellite orbits, \citet{2025arXiv250312612M} estimate the total mass of the MW within 240~kpc to be $M_\mathrm{MW} = ( 0.79 \pm 0.23 ) \times 10^{12}$~\Msun{}.
Similarly, \citet{2025arXiv250312612M} measure the mass of the M\,31 satellite system to be $M_\mathrm{M31} = ( 1.55 \pm 0.34 ) \times 10^{12}$~\Msun{} within 300~kpc.

Taking the mass ratio to be 2.0 from the last measurements, we can fix the position of the center of mass of the Local Group and apply the Hubble flow model.
Seven of the eleven galaxies provide consistent estimates of the total mass of the Local Group with a mean of $M_\mathrm{LG} = (2.43 \pm 0.21) \times 10^{12}$~\Msun{} and a surprisingly small spread of $\sigma_M = 0.57 \times 10^{12}$~\Msun{}.
An important point is the good agreement between this mass estimate and the sum of the individual MW and M\,31 masses, $M_\mathrm{MW+M31} = (2.34 \pm 0.41) \times 10^{12}$~\Msun{}, with both values based on completely independent approaches.
This indicates that the method is effective and produces a result close to the expected mass.
Moreover, it follows that essentially the entire mass of the Local Group is concentrated around the two main galaxies.

\begin{figure}
\centering
\includegraphics[width=\linewidth, clip]{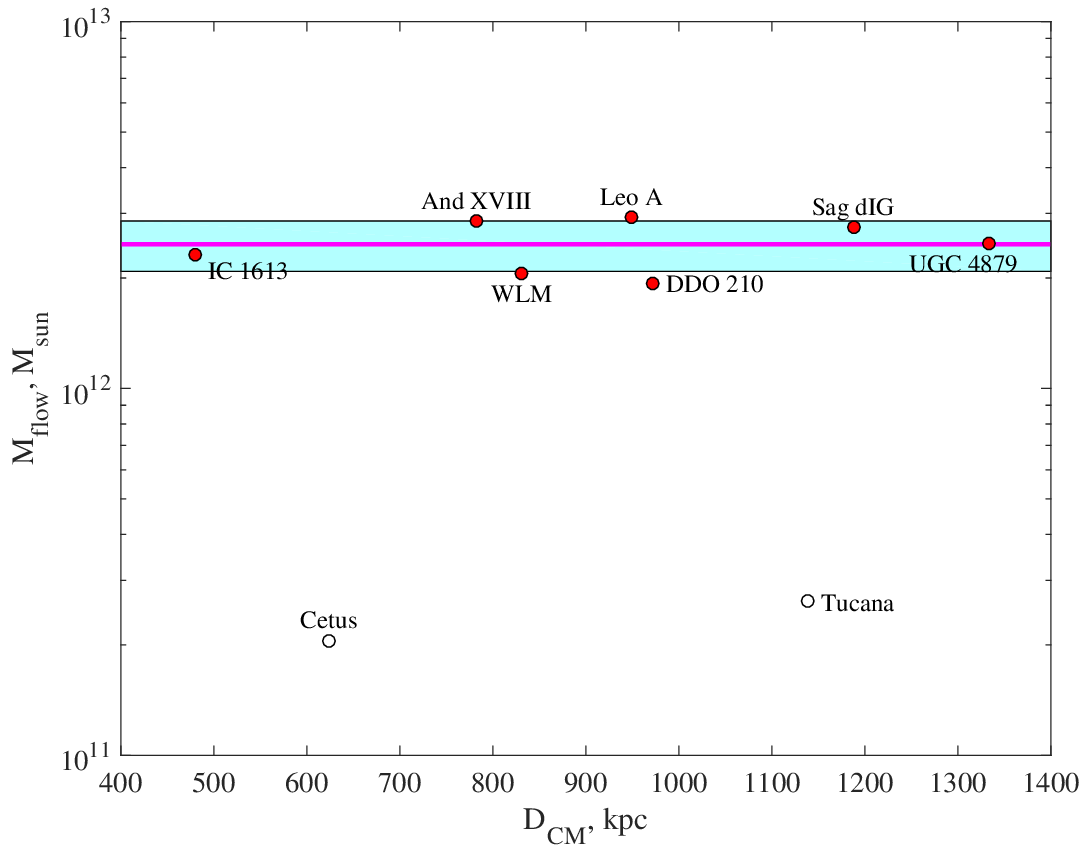}
\caption{
The total mass of the Local Group from the Hubble flow model versus the distance to the center of mass of the Local Group, which minimizes the spread of the mass estimates.
}
\label{fig:Mass}
\end{figure}

A natural extension of the analysis is to determine the position of the barycenter that minimizes the observed spread in total mass estimates.
Given the previous result that the sum of the individual masses of the MW and M\,31, inferred from the kinematics of the satellites, coincides with the mass of the Local Group derived from the motion of its outer members, we can be confident that the center of mass of the system lies on the line connecting the MW and M\,31.
By varying the MW-to-M\,31 mass ratio---and consequently the position of the barycenter---we can significantly reduce the spread in the Local Group mass estimates, by a factor of 1.5 to $\sigma_M = 0.39 \times 10^{12}$~\Msun{}.
In this case, the total mass is $M_\mathrm{LG} = (2.47 \pm 0.15) \times 10^{12}$~\Msun{}.
The result is shown in Fig.~\ref{fig:Mass}.

The minimum scatter is achieved for the MW-to-M\,31 mass ratio of $M_\mathrm{MW}/M_\mathrm{M31} = 0.74 \pm 0.10$, placing the barycenter $447\pm26$~kpc from the MW center and implying the MW velocity of $62.6\pm2.6$~\kms{} in the direction of the barycenter.
This allows us to estimate individual masses $M_\mathrm{MW} = (1.06\pm0.11) \times 10^{12}$~\Msun{} and $M_\mathrm{M31}=(1.42\pm0.12) \times 10^{12}$~\Msun{}.
The results on the mass estimates of the Local Group and the MW-to-M\,31 mass ratio are discussed in more detail in Section~\ref{sec:Discussion}.

We do not detect any statistically significant trend of the mass with distance from the center of mass of the Local Group.
This indicates that over a wide range of distances from 400~kpc to 1.4~Mpc, the total mass of the system is confined within a region of the order of 400~kpc, effectively inside the virial zones around the MW and M\,31.

\section{Comparison with HESTIA simulations}

We compare with \LCDM{} simulations of the Local Group, known as the HESTIA simulations~\citep{2020MNRAS.498.2968L}. The HESTIA runs are constrained Magneto-Hydro-dynamic simulations of the Local Group performed in a constrained environment which mimics the observed one quite closely. The simulations have been run with the AURIGA \citep{2017MNRAS.467..179G} model of galaxy formation. The initial conditions were created by mapping the $z=0$ gravitational environment via a Wiener Filter reconstruction from CosmicFlows--2 survey of peculiar velocities~\citep{2013AJ....146...86T}, via a reverse Zeldovich approximation \citep{2013MNRAS.430..888D}.  

The simulations result in a local group constituting of two spiral galaxies whose mass (i.e. 0.8--$1.2 \times 10^{12}$~\Msun{}), mass ratio ($\sim$~1:1) and separation ($\sim$ 700--800~kpc) are consistent with measured observational properties. Furthermore, properties such as circular velocity, size and star formation history are also all within the observed values. Further afield, the location of a Virgo-like cluster, as well as other cosmographic features (local void, sheet and filament) are all natural outcomes of the constraining technique. Since these are high resolution simulations (achieving a mass resolution of $m_\mathrm{dm}=1.2 \times 10^6$ and $m_\mathrm{gas} = 1.8 \times 10^5$~\Msun{} for dark matter and gas, respectively) the Local Group (and beyond) is also populated by an entourage of low mass dwarf and satellite galaxies.

\begin{figure}
\centering
\includegraphics[width=\linewidth, clip]{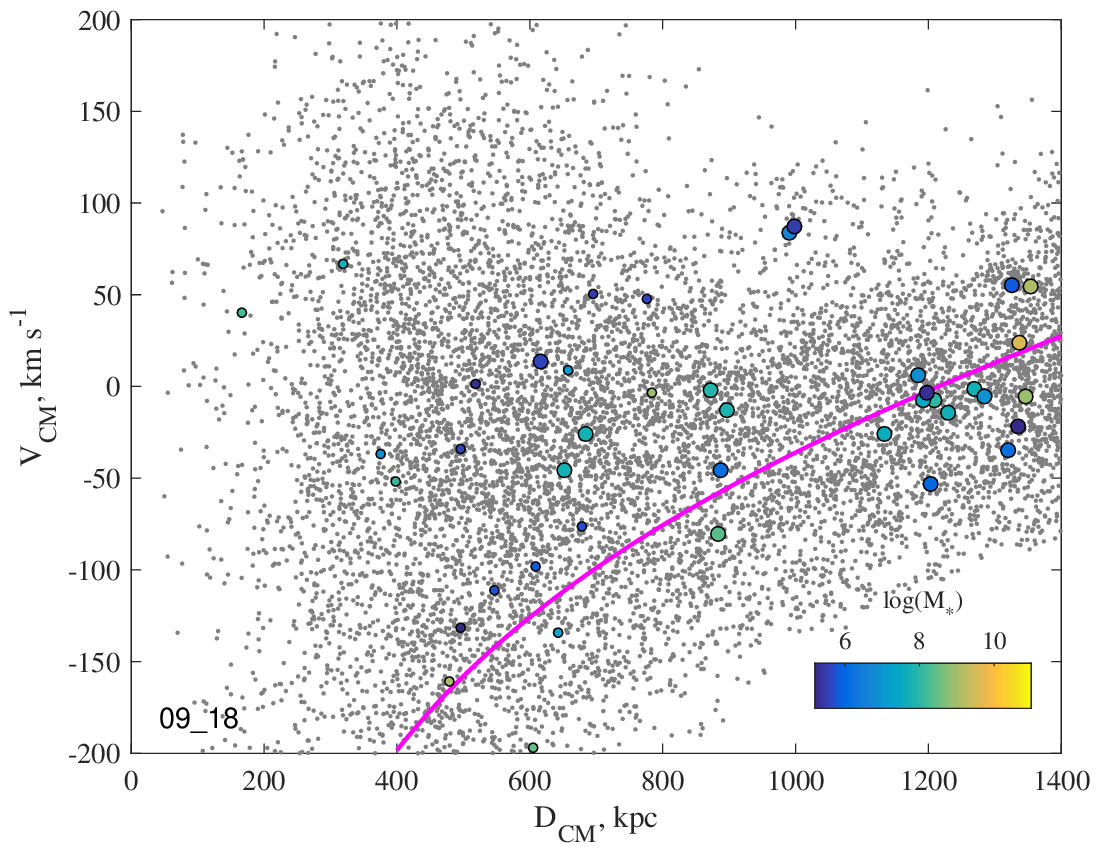}
\caption{
The Hubble flow in the neighborhood of the Local Group analog in the high-resolution HESTIA simulations 09\_18.  
Colored circles indicate halos with stellar particles, $M_*\ge1.5\times10^5$~\Msun{}, but only outside the virial zones of 300~kpc around of two most massive halos.
The large circles correspond to distant members $D\ge450$~kpc away from the MW and M\,31 analogs, while the small circles indicate closer objects.
}
\label{fig:SimulatedHubbleFlow}
\end{figure}

\begin{figure}
\centering
\includegraphics[width=\linewidth, clip]{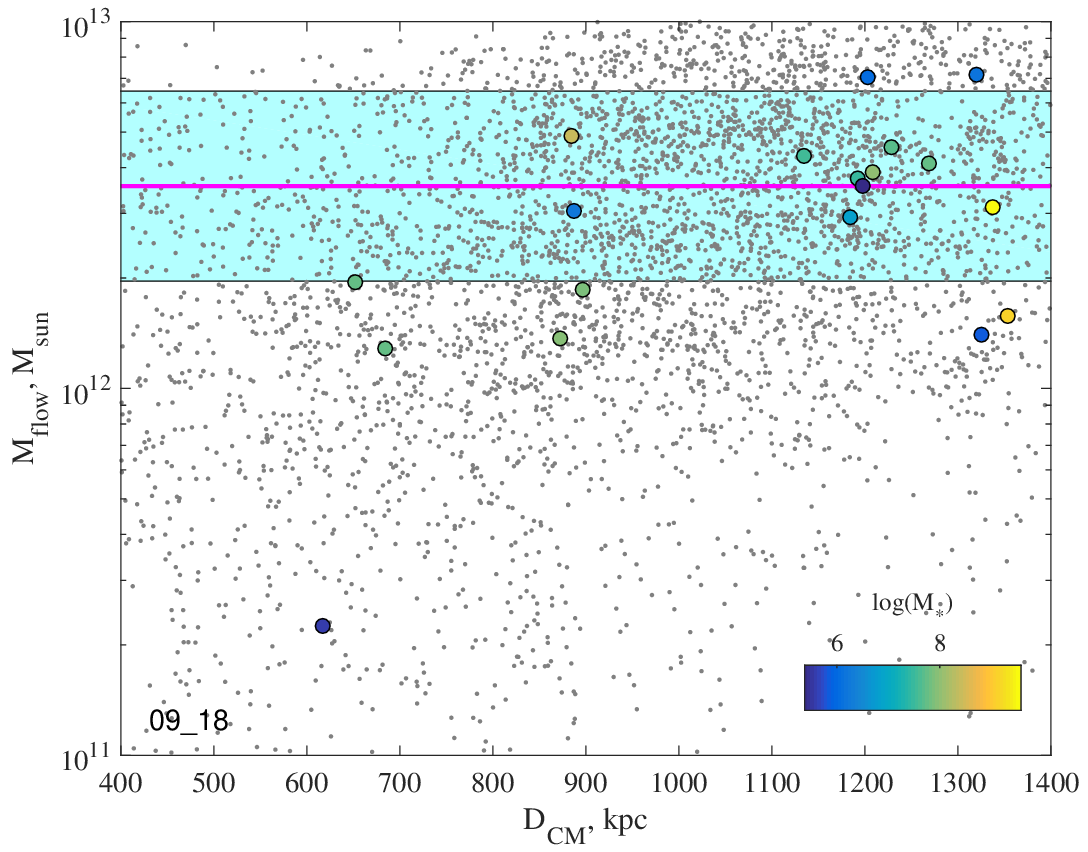}
\caption{
Estimation of the mass of the system from the Hubble flow, as in Fig.~\ref{fig:Mass}, but using the HESTIA simulation data.
}
\label{fig:SimulatedMass}
\end{figure}

In Figure~\ref{fig:SimulatedHubbleFlow}, we show an example of the halo velocity-distance diagram in the neighborhood of the Local Group analog obtained in the high-resolution HESTIA simulation 09\_18. 
This plot mimics the real observations when the observer is positioned at the center of the second most massive halo in the group, with the line-of-sight halo velocities and distances recalculated to the center of mass of the system, similar to Fig.~\ref{fig:HubbleFlow}.
The halos with stars, $M_*\ge1.5\times10^5$, are shown by color circles only outside the virial zones of 300~kpc around of two most massive halos.
The large circles show halos at $D\ge450$~kpc from the central halos. 
They correspond to the real galaxies used to estimate the Local Group mass.
The small colored circles represent ``satellites'' in the vicinity of the two central galaxies, $300\le D \le450$~kpc.
The virial zones of the two massive halos overlap and are centered at a distance of about 400~kpc relative to the center of mass.
Just outside the virial zones, a well-pronounced flow of objects is visible as they fall into the group.
Above this flow there are a number of outliers representing, among others, the backsplash objects.
Star-forming halos track the general behavior of all halos well.
It can be seen that the simulations reproduce the observed velocity-distance diagram quite well.
The main difference between the simulations and the Local Group is in the significantly larger velocity spread of the infalling galaxies.
The typical value in the simulations is about 70~\kms{}, compared to 15~\kms{} for galaxies following the cold Hubble flow outside the virial zones in the Local Group.

\begin{table}
\centering
\caption{
Comparison of mass estimations using the Hubble flow model with the HESTIA simulations
}
{\small
\begin{tabular}{cccrcc}
\hline\hline
Set    & $M_1+M_2$ & $M(r<1)$ & & $\langle M_\mathrm{flow} \rangle$ & $\sigma$\\
\cline{2-3} \cline{5-6}
       & \multicolumn{2}{c}{$\times 10^{12}$~\Msun{}} & & \multicolumn{2}{c}{$\times 10^{12}$~\Msun{}} \\
\hline

09\_18 &  2.76      & 3.33                  & 4980 & $3.27 \pm 0.04$ & 2.62 \\
\multicolumn{3}{r}{$M_*\ge1.5\times10^5$}   &   21 & $3.51 \pm 0.41$ & 1.95 \\

17\_11 &  2.89      & 3.33                  & 4902 & $3.54 \pm 0.04$ & 3.42 \\
\multicolumn{3}{r}{$M_*\ge1.5\times10^5$}   &   15 & $4.87 \pm 1.00$ & 4.37 \\

37\_11 &  1.40      & 2.08                  & 4028 & $2.72 \pm 0.03$ & 2.18 \\
\multicolumn{3}{r}{$M_*\ge1.5\times10^5$}   &   27 & $2.36 \pm 0.29$ & 1.68 \\

\hline\hline
\end{tabular}
}
\label{tab:HESTIA}
\end{table}

\section{Discussion}
\label{sec:Discussion}

First of all, it is important to highlight the existence of two distinct populations of galaxies located outside the virial zones of the Milky Way and the Andromeda Galaxy.
Two-thirds of the known galaxies follow a regular Hubble flow relative to the barycenter of the Local Group.
The line-of-sight velocity scatter around this flow is only 15~\kms{}, which may be partially attributed to uncertainties in distance measurements.
The remaining galaxies exhibit different kinematics, as discussed below.

The extremely cold Hubble flow in the nearby Universe was pointed out by \citet{2001Afz....44....1K,  2002A&A...389..812K}, who discovered a line-of-sight dispersion of 25--30~\kms{} in the expansion of isolated nearby galaxies up to $\sim$3~Mpc, based on precise the Tip of the Red Giant Branch (TRGB) distances.
\citet{2001A&A...368L..17E} found the velocity dispersion of 38~\kms{} in the distance range 1--8~Mpc using Cepheid distances.
Such a regular and cold Hubble expansion in the face of the highly nonuniform environment in the nearby Universe seemed mysterious.
Attention to this problem goes back as far as \citet{1972ApJ...172..253S}.
\citet{2001A&A...378..729B} suggested that this enigma is explained by adiabatic cooling of the chaotic motions of galaxies in regions dominated by dark energy.
In contrast, the insignificant contribution of $\Lambda$ to the formation of the density profile and velocity field around galaxy clusters was pointed out as early as \citet{1991MNRAS.251..128L}.
Using constrained cosmological simulations, \citet{2008MNRAS.386..390H} showed that dark energy does not manifest itself in any way in the dynamics of the local Universe, and the dispersion of $\lesssim60$~\kms{} in the range from 0.75 to 2 or 3~Mpc is found in more than half of the cases. 
Thus, they conclude that the problem of the cold Hubble flow is not related to dark energy, but rather to the mean matter density around objects.

Our sample size is rather small: only 7 of the 11 galaxies considered follow the cold Hubble flow.
However, it is worth noting that 4 of the 6 satellites located near the virial zones of MW and M\,31---excluded from the analysis because of their proximity to these massive galaxies---also closely match this cold flow.
This provides additional support for the reality of the effect.

It is important to emphasize that we observe the cold Hubble flow with a velocity dispersion of only 15~\kms{} within the zero-velocity sphere of the Local Group, at distances ranging from 400 to 1400~kpc.
If this is not a statistical anomaly, such low chaotic motion is surprising, even despite the fact that the nearest massive neighbors lie 3--4~Mpc away from the Local Group.
Current cosmological models predict a steady inflow of matter into galaxy groups and clusters.
But, even in constrained simulations such as HESTIA, which reproduce the Local Group environment reasonably well, the resulting flows are not as cold.

Despite the above, not all galaxies conform to the Hubble flow.
The positions and velocities of the four galaxies cannot be described by the model under any reasonable parameters. 
Apparently, this may indicate that they are not at the stage of the first infall into the Local Group and have experienced interactions with its members before.

Tucana is an isolated dSph galaxy on the border of the Local Group.
\citet{2021MNRAS.502.1623M} found that it does not show any sign of residual star formation in the last 2~Gyr, which differs it significantly from other isolated dSph galaxies, KKR\,25, KK\,258, and KKs\,3, as well as And\,XVIII, which is located within the Local Group but well outside the virial zone of M\,31.
Tucana's star formation is typical of dSph satellites of giant galaxies that have lost their gas through ram pressure stripping and tidal interaction with a massive galaxy.
Note that And\,XVIII follows the Hubble flow perfectly.

Cetus, like Tucana, is an isolated dSph galaxy located within the Local Group. It does not follow the Local Group morphology-density relation. 
\citet{2018A&A...618A.122T} note tidal tail-like structures in the outer part of the galaxy, although their presence is difficult to reconcile with the timescales of a possible passage around M\,31, and the orbit of Cetus through the Local Group indicates that it is at the apocenter~\citep{2007MNRAS.375.1364L}.
\citet{2015ApJ...811L..18G} suggested that the line-of-sight velocities of Tucana and Cetus are compatible with the assumption that they were close to the barycenter of the Local Group at early times, probably when they just assembled.
Likewise, \citet{2012MNRAS.426.1808T} suggested that Tucana, Cetus, NGC\,6822 are possibly backsplash galaxies, based on their current observed positions and radial velocities.
Data, presented in the study of \citet{2003ApJ...590L..17K}, support the scenario that a passage of the northwest \ion{H}{I} cloud produced the ``tidal arm'' to the southeast of NCG~6822 and triggered the recent star formation activity found in the entire galaxy. 
At the same time, \citet{2021MNRAS.501.2363M} find that NGC\,6822 is highly likely to be isolated and on its first infall, based on orbital properties derived using Gaia Data Release~2 proper motions of the brightest stars.

\citet{2009MNRAS.400.2054K} studied the stellar component of the very faint outer regions of the Pegasus dwarf (Peg~DIG), as well as its \ion{H}{I} component. 
The authors rule out a possible tidal origin or a ram pressure stripping scenario, and propose that the Pegasus dwarf is on its first infall in the Local group since it does not appear to be disturbed by interactions with other galaxies.
However, it is also worth to note that, according to the studies mentioned above, NGC\,6822 and Peg~DIG are dark-matter dominated objects.

Thus, the four outliers, Cetus, Tucana, Peg~DIG and NGC\,6822 are distinct from other studied external galaxies, and the reasons may be suspected in their evolution in the Local Group (backsplash galaxies) as well as in their assembly and star formation details (dark matter-dominated objects).

The arguments presented above give us reason to believe that we obtain a reliable estimate of the total mass of the Local Group $M_\mathrm{LG} = (2.47 \pm 0.15) \times 10^{12}$~\Msun{} by studying the kinematics of the peripheral members of the Local Group.

The idea of using deviations from the ideal linear Hubble flow to estimate the total mass of the Local Group dates back to \citet{1981Obs...101..111L} and \citet{1986ApJ...307....1S}.
The radius of the zero-velocity sphere, determined from the recession of nearby galaxies, has been used many times for this purpose.
\citet{2009MNRAS.393.1265K} measured $R_0 = (0.96 \pm 0.03)$~Mpc, which corresponds to a total mass of $M_\mathrm{LG} = (1.9 \pm 0.2) \times 10^{12}$~\Msun{}.
\citet{2018A&A...609A..11K} obtained even smaller values of $R_0 = (0.91 \pm 0.05)$~Mpc and $M_\mathrm{LG} = (1.5 \pm 0.2) \times 10^{12}$~\Msun{}, and noted the paradoxical situation when the total mass turns out to be less than the sum of the virial masses of MW and M\,31.
In our work, we account for the influence of dark energy more carefully and use the analytical model of the Hubble flow~\citep{2020PhRvD.102h3529B}, which provides a larger and more precise estimate of the total mass of the Local Group.
This, in turn, resolves the aforementioned paradox.
\citet{2024PhLB..85839033B} refine the definition of zero-acceleration and zero-velocity surfaces for a general mass distribution on a cosmological background governed by a cosmological constant.
They estimated the total mass of the Local Group to be $M_\mathrm{LG} = (2.47 \pm 0.08) \times 10^{12}$~\Msun{}, which matches our value perfectly.

Another classic approach to weighing the Local Group is the so-called timing argument, proposed by \citet{1959ApJ...130..705K}.
It is based on the assumption that the observed configuration of the Milky Way and the Andromeda Galaxy is the result of their first encounter while moving along radial Keplerian orbits during the lifetime of the Universe.
This method usually produces a significantly larger mass of the Local Group of about $5\times10^{12}$~\Msun{} compared to other methods~\citep{2023MNRAS.521.4863S}.
So, \citet{2022ApJ...928L...5B} estimate the total mass of $(3.7\pm0.5)\times10^{12}$~\Msun{} taking into account the proper motion of M\,31 provided by the Gaia mission.
However, careful consideration of the influence of the Large Magellanic Cloud reduces this value to $(2.3\pm0.7)\times10^{12}$~\Msun{}~\citep{2024A&A...689L...1B}, making the timing argument comparable with other estimates and, in turn, agreeing within error with our value.

We would like to emphasize two important results of our work:
1) the absence of a significant trend in the estimate of the total mass of the Local Group in the distance range from 400 to 1400~kpc;
2) the equality of the Local Group mass of $M_\mathrm{LG} = (2.47 \pm 0.15) \times 10^{12}$~\Msun{} estimated from the Hubble flow to the sum of the individual masses of the Milky Way and the Andromeda Galaxy, $M_\mathrm{MW+M31} = (2.34 \pm 0.41) \times 10^{12}$~\Msun{}~\citep{2025arXiv250312612M}, measured from the kinematics of their satellites.
This indicates that almost all of the mass of the Local Group is concentrated within the virial zones of its two main galaxies.
\citet{2005AJ....129..178K, 2009MNRAS.393.1265K} previously highlighted the equality of the virial masses of galaxy groups on scales of 200--300~kpc with their total masses inside a zero-velocity sphere of about 1~Mpc in radius.
This is probably also true for larger structures, such as the Virgo cluster~\citep{2010MNRAS.405.1075K}.

The fact that all the mass is concentrated around MW and M\,31 allows us to determine their mass ratio of $M_\mathrm{MW}/M_\mathrm{M31} = 0.74 \pm 0.10$ by finding the barycenter of the Local Group to minimize the spread of the mass estimates from the Hubble flow model.
It corresponds to the individual masses of $M_\mathrm{MW} = (1.06 \pm 0.11) \times 10^{12}$~\Msun{} and $M_\mathrm{M31}=(1.42 \pm 0.12) \times 10^{12}$~\Msun{} for the Milky Way and the Andromeda Galaxy, respectively.
This MW mass is somewhat greater than the mass estimated from the kinematics of its satellites excluding Leo\,I, $(0.79\pm0.23)\times10^{12}$~\Msun{}~\citep{2025arXiv250312612M}, although it remains in agreement.
However, if Leo\,I is included in the consideration, the MW mass increases to $(1.07\pm0.30)\times10^{12}$~\Msun{}, which perfectly matches the estimate from the barycenter position.
This is also consistent with numerous measurements of the MW mass within 200~kpc made by other authors and collected in recent reviews.
The compilation by \citet{2020SCPMA..6309801W} gives an average of $\langle M_\mathrm{MW} \rangle = 1.07\times10^{12}$~\Msun{} with a spread of $\sigma_\mathrm{MW} = 0.30\times10^{12}$~\Msun{},
and \citet{2023ARep...67..812B} point out a mean of $\langle M_\mathrm{MW} \rangle = 0.88\times10^{12}$~\Msun{} with a dispersion of $\sigma_\mathrm{MW} = 0.24\times 10^{12}$~\Msun{}.
The M\,31 mass estimated from the barycenter position also agrees well with the kinematics of its satellites within 300~kpc, both from the literature, $\langle M_\mathrm{M31} \rangle = 1.61 \times 10^{12}$~\Msun{}, $\sigma_\mathrm{M31} = 0.24 \times 10^{12}$~\Msun{}~\citep[see review by][]{2023arXiv230503293B}, and from our measurements $M_\mathrm{M31} = (1.55 \pm 0.34) \times 10^{12}$~$M_{\odot}$~\citep{2025arXiv250312612M}.

The barycenter of the Local Group, determined from the seven dwarfs, is located at a distance of $447\pm26$~kpc from the MW to the M\,31 centers, which corresponds to a relative distance of $0.57\pm0.03$, given the separation of 780.3~kpc between the two galaxies.
This result is in excellent agreement with the value of $0.55\pm0.05$ reported by \citet{2009MNRAS.393.1265K}, who used 30 neighboring galaxies with distances ranging from 0.7 to 3.0~Mpc.
Taking into account the approach velocity between the MW and M\,31 of $-109\pm2$~\kms{}, we find that the Milky Way is moving at a speed of $62.6\pm2.6$~\kms{} toward the barycenter of the Local Group at $(l,b) = (+121.7^\circ \pm 2.2^\circ, -21.5^\circ \pm 1.5^\circ)$.
Combining this with the motion of the Sun in the Galaxy, we obtain the solar vector relative to the barycenter of the Local Group equal to $(V_x,V_y,V_z) = ( -21.1 \pm 3.5 , +300.3 \pm 3.4 , -14.3 \pm 1.4 )$ in the Galactic coordinates, which in the sky corresponds to $(l,b,V) = ( +94.0^\circ \pm 0.7^\circ ,  -2.7^\circ \pm 0.3^\circ , 301.3 \pm 3.3~\kms{})$.
This is in excellent agreement with the apex of the Sun relative to the centroid of the Local Group, $( +93^\circ \pm 2^\circ , -4^\circ \pm 2^\circ , 316 \pm 5~\kms{})$, found by \citet{1996AJ....111..794K}.

\section{Conclusions}

We analyze the motions of 11 peripheral members of the Local Group at distances from 400 to 1400~kpc from its center.
The positions and velocities of 7 of them are perfectly described by an analytical model of the Hubble flow around a spherical overdensity in the standard flat \LCDM{} universe~\citep{2020PhRvD.102h3529B}.
The remaining 4 galaxies have most likely been influenced earlier by other members of the Local Group and cannot be considered as free-falling particles.
In the following, we summarize our conclusions.

\begin{itemize}

\item 
Despite the dumbbell-shaped structure of the Local Group, the spherically symmetric Hubble flow model~\citep{2020PhRvD.102h3529B} relative to the barycenter of the Local Group gives a very good description of the velocity field inside its zero-velocity sphere down to the boundaries of the virial zones around the Milky Way and the Andromeda Galaxy.

\item 
The Hubble flow in the vicinity of the Local Group from 400 to 1400~kpc is extremely cold with a line-of-sight velocity dispersion of only 15~\kms{}.
This is actually an upper limit, as distance errors should contribute significantly.

\item 
Constrained cosmological \LCDM{} simulations predict a much larger scatter of the velocity field, on the order of 70~\kms{}, outside the virial radii of the MW and M\,31.

\item 
Most mass estimates derived from the velocity and position of individual galaxies relative to the Local Group barycenter agree well with each other with extremely small scatter of $\sigma_M = 0.39 \times 10^{12}$~\Msun{}.
This allows us to estimate the total mass of the Local Group as $M_\mathrm{LG} = (2.47 \pm 0.15) \times 10^{12}$~\Msun{}.

\item 
There is no statistically significant change in the mass of the Local Group in the distance range from 400 to 1400~kpc.

\item 
The Hubble flow model is in remarkable agreement with the sum of the total masses of our Galaxy and the Andromeda Galaxy, estimated from the kinematics of their satellites.
From this we conclude that almost the entire mass of the Local Group is concentrated within 400~kpc around these two giant spirals, in fact within their virial radii.

\item 
The barycenter of the Local Group, determined from the condition of minimizing the spread of mass estimates, corresponds to the mass ratio between the Milky Way and the Andromeda Galaxy equal to $M_\mathrm{MW}/M_\mathrm{M31} = 0.74\pm0.10$.
Our Galaxy is moving toward the barycenter at a speed of $62.6\pm2.6$~\kms{}.
Thus, the solar apex relative to the barycenter of the Local Group is equal to $(l,b,V) = ( +94.0^\circ \pm 0.7^\circ ,  -2.7^\circ \pm 0.3^\circ , 301 \pm 3~\kms{})$ in the Galactic coordinates.

\end{itemize}

\begin{acknowledgements}
This work was supported by the Russian Science Foundation grant \textnumero~24--12--00277. 
\end{acknowledgements}

\bibliographystyle{aa}
\bibliography{ref}

\end{document}